\documentclass[twocolumn,prl,showpacs,preprintnumbers,amsmath,amssymb,superscriptaddress]{revtex4-1}

\usepackage{graphicx}               % inclusion of graphics files
\usepackage{dcolumn}                % alignment of table columns on decimal point
\usepackage{hyperref}

\begin{document}

%\preprint{APS/123-QED}
\preprint{Accepted to \prl}

\title{Collisional Thermalization of Hydrogen and Helium in Solar Wind Plasma}

\author{B.~A.~Maruca}
\email{bmaruca@ssl.berkeley.edu}
\affiliation{Space Sciences Laboratory, University of California, 7 Gauss Way, Berkeley, CA 94720, USA}

\author{S.~D.~Bale}
\affiliation{Space Sciences Laboratory, University of California, 7 Gauss Way, Berkeley, CA 94720, USA}
\affiliation{Department of Physics, University of California, LeConte Hall, Berkeley, CA 94720, USA}

\author{L.~Sorriso-Valvo}
\affiliation{IPCF-CNR, U.~O.~Cosenza, Ponte P.~Bucci, Cubo 31C, 87036 Rende, Italy}
\affiliation{Space Sciences Laboratory, University of California, 7 Gauss Way, Berkeley, CA 94720, USA}

\author{J.~C.~Kasper}
\affiliation{Department of Atmospheric, Oceanic and Space Sciences, University of Michigan, Ann Arbor, MI 48109, USA}
\affiliation{Harvard-Smithsonian Center for Astrophysics, Cambridge, MA, USA}

\author{M.~L.~Stevens}
\affiliation{Harvard-Smithsonian Center for Astrophysics, Cambridge, MA, USA}

\date{\today}


% 96.50.Ci  Solar wind plasma; sources of solar wind 
% 52.20.Hv  Atomic, molecular, ion, and heavy-particle collisions
% 52.25.Kn  Thermodynamics of plasmas

\begin{abstract}
In situ observations of the solar wind frequently show the temperature of $\alpha$-particles (fully ionized helium), $T_\alpha$, to significantly differ from that of protons (ionized hydrogen), $T_p$.  Many heating processes in the plasma act preferentially on $\alpha$-particles, even as collisions among ions act to gradually establish thermal equilibrium.  Measurements from the \textit{Wind} spacecraft's Faraday cups reveal that, at $r=1.0\ \textrm{AU}$ from the Sun, the observed values of the $\alpha$-proton temperature ratio, $\theta_{\alpha p} \equiv T_\alpha\,/\,T_p$ has a complex, bimodal distribution.  This study applied a simple model for the radial evolution of $\theta_{\alpha p}$ to these data to compute expected values of $\theta_{\alpha p}$ at $r=0.1\ \textrm{AU}$.  These inferred $\theta_{\alpha p}$-values have no trace of the bimodality seen in the $\theta_{\alpha p}$-values measured at $r=1.0\ \textrm{AU}$ but are instead consistent with the actions of the known mechanisms for $\alpha$-particle preferential heating.  This result underscores the importance of collisional processes in the dynamics of the solar wind and suggests that similar mechanisms may lead to preferential $\alpha$-particle heating in both slow and fast wind.
\end{abstract}

\maketitle

%%% Introduction, motivation, and observations

The solar wind is the highly-ionized, magnetized plasma that flows supersonically from the Sun's corona into deep space.  Though its composition varies considerably, protons (ionized hydrogen) and $\alpha$-particles (fully-ionized helium) constitute the vast majority of ions \cite{marsch82a,kasper12}.  The $\alpha$-proton relative abundance, $n_{\alpha}\,/\,n_p$, where $n_j$ is number density (with $j=p$ for protons and $\alpha$ for $\alpha$-particles), rarely exceeds $20.\%$ and is usually about $4.\%$.

The solar wind's low density and high temperature ensure that collisions among its constituent particles only affect plasma dynamics on relatively long timescales.  Expansion, wave-particle interactions, and turbulence can influence the solar wind on much shorter timescales and thus frequently cause deviations from thermodynamic equilibrium \cite{marsch06,schekochihin09}.  Particle species often have different bulk velocities and temperatures.  Additionally, distinct temperatures, $T_{\perp j}$ and $T_{\parallel j}$, can develop along the directions perpendicular and parallel to the background magnetic field.  The (\textit{scalar}) \textit{temperature} of a species is then the weighted average of its \textit{component temperatures}: i.e.,
\begin{equation} \label{eqn:tj}
T_j \equiv \left( 2\,T_{\perp j} + T_{\parallel j} \right)/\,3\,.
\end{equation}

This Letter focuses specifically on one non-equilibrium feature of solar wind plasma: the unequal temperatures of protons and $\alpha$-particles.  This phenomenon can be quantified by the $\alpha$-proton relative temperature:\begin{equation} \label{eqn:theta}
\theta_{\alpha p} \equiv T_\alpha\,/\,T_p\,.
\end{equation}
While the $\alpha$-proton relative temperature components, 
\begin{equation} \label{eqn:theta:comp}
\theta_{\perp\alpha p} \equiv T_{\perp\alpha}\,/\,T_{\perp p}\ \ \textrm{and}\ \ \theta_{\parallel\alpha p} \equiv T_{\parallel\alpha}\,/\,T_{\parallel p}\,,
\end{equation}
are considered to some extent herein, the proceeding analysis principally uses the distribution of observed $\theta_{\alpha p}$-values to elucidate the effects of particle collisions and other processes on ion temperatures.

\begin{figure}
\includegraphics[width=3in]{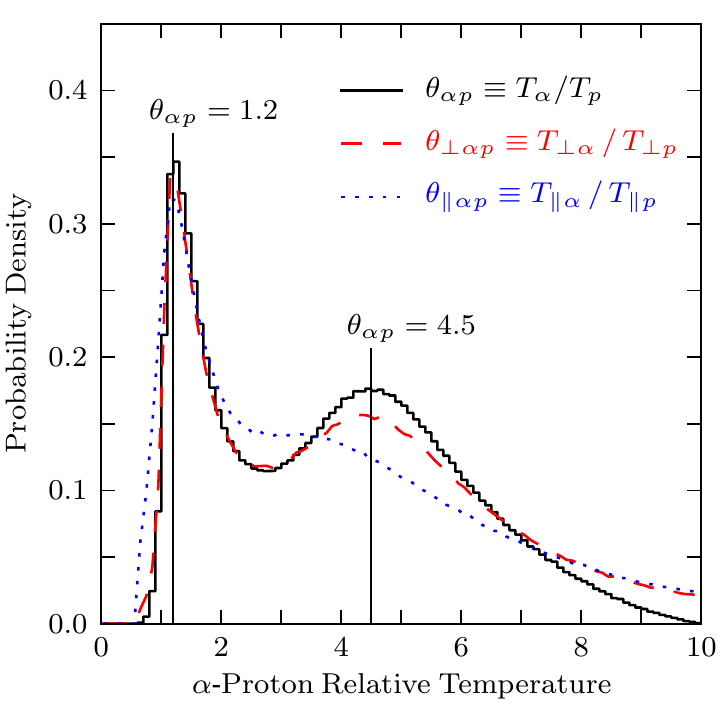}
\caption{\label{fig:meas}Distribution of $\theta_{\alpha p}$-values observed with the \textit{Wind} spacecraft.  The distributions of $\theta_{\perp\alpha p}$ and $\theta_{\parallel\alpha p}$ are also shown (red, dashed and blue, dotted curves, respectively).}
\end{figure}

The black, solid histogram in Figure \ref{fig:meas} shows the probability distribution of $\theta_{\alpha p}$-values observed in the solar wind at $r = 1\ \textrm{AU}$ from the Sun.  The dataset used for this figure (and for all the analysis described in this Letter) was derived from in situ measurements of solar wind ions from the {\it Wind} spacecraft's Faraday cups \cite{ogilvie95}.  This instrument produces an \textit{ion spectrum}: a distribution of ion speeds projected along various axes.  Proton and $\alpha$-particle bulk parameters are inferred from each spectrum by fitting a bi-Maxwellian model for the core of each species' velocity distribution function \cite{kasper06,maruca13}.  Measurements of the local magnetic field \cite{lepping95} are used to separate perpendicular and parallel temperature components \cite{kasper06,maruca13}.

The dataset used for this study was compiled from $2.1$-million ion spectra that were processed in this way.  Initially, $4.8$-million spectra (i.e., all spectra from the spacecraft's launch in late-1994 through mid-2010) were processed, but the final dataset only included spectra which met two criteria \cite{maruca11}.  First, a spectrum needed to have been measured at a time when \textit{Wind} was well outside Earth's bow shock (i.e., actually in the solar wind).  Early in its mission, the spacecraft spent considerable time exploring Earth's magnetosphere.  Second, the fit results had to be of high quality.  Often, the failure of this latter criterion resulted from an $\alpha$-particle signal that was weak, out of the instruments energy range, or confused with the proton signal (i.e., low-$n_\alpha$, high-$v_\alpha$, and high-$T_p$, respectively).  While this did produce some bias in the final dataset, a wide range of solar wind conditions is still well represented \cite{maruca11,maruca12}.

While high-quality bulk-parameter values can be inferred from Faraday cup data, actually quantifying the uncertainty in an individual value is non-trivial \cite{kasper06}.  Nevertheless, statistical analyses can be used to gauge the overall uncertainty in the values of a given parameter.  For example, the ratio of a parameter's standard deviation to it's mean can be computed over many short intervals; then, the median of this ratio gives an upper bound on the ``typical,'' random error for that parameter.  Applied to the dataset used in this study, this method indicates relative uncertainties of $7.6\%$ in $T_p$ and $15.\%$ in $T_\alpha$.  This is consistent with the $8.\%$ uncertainty in $T_p$ that derived for a similar \textit{Wind} Faraday cup dataset that was established via an independent method \cite{kasper06}.

The distribution of observed $\theta_{\alpha p}$-values in Figure \ref{fig:meas} has two distinct peaks (i.e., is bimodal).  By fitting a Gaussian to the crest of each peak, these modes were quantified as $\theta_{\alpha p} = 1.2$ and $4.5$.  The former corresponds to plasma with protons and $\alpha$-particles nearly in thermal equilibrium (i.e., $\theta_{\alpha p} = 1$).  The latter indicates plasma in which the $\alpha$-particles have been preferentially heating relative to the protons.  Since an $\alpha$-particle's mass is about four-times a proton's, this heating was approximately mass-proportional (i.e., produced approximately equal thermal speeds).  Numerous studies have observed such mass-proportional temperatures between many different ion species in the solar wind \cite{schmidt80,marsch82a,vonsteiger95,hefti98}.  Proposed mechanisms for the preferential heating of $\alpha$-particles and other heavy ions include include drift instabilities \cite{gary00a,gary00b}, low-frequency Alfv\'{e}n-wave turbulence \cite{chandran10}, and resonant absorption of ion-cyclotron waves \cite{isenberg07,isenberg09,kasper13}.

All of these preferential-heating mechanisms would principally increase perpendicular temperature, and the distribution of observed $\theta_{\perp\alpha p}$- and $\theta_{\parallel\alpha p}$-values (plotted in Figure \ref{fig:meas} with colored, broken curves) offer some evidence of this.  While, overall, these distributions resemble that for $\theta_{\alpha p}$, the second peak in the $\theta_{\parallel\alpha p}$-curve occurs at a markedly lower value and is less defined.  This is consistent with the preferential heating of $\alpha$-particles being principally perpendicular; high $\theta_{\parallel\alpha p}$-values would then result indirectly via $\alpha$-particle isotropization from particle collisions and/or instabilities \cite{maruca12}.  Even so, for $\alpha$-proton relative temperatures $\gtrsim 7$, both the $\theta_{\perp\alpha p}$ and $\theta_{\parallel\alpha p}$ distributions exceed that of $\theta_{\alpha p}$.  This would indicate that such extreme values of $\theta_{\perp\alpha p}$ and $\theta_{\parallel\alpha p}$ do not occur concurrently, which may then suggest that a mechanism for extreme parallel $\alpha$-particle heating exists.

%%% Collisional age

One paradigm for exploring the origins of the $\theta_{\alpha p}$-distribution's bimodality is solar wind speed, which, for convenience, is often (and herein) taken to be $v_{rp}$, the radial component of proton velocity.  Wind of different speeds originates from different parts of the Sun and via different mechanisms \cite[and references therein]{marsch06}, and many solar wind properties (e.g., composition, charge state, temperature, and turbulence) trend closely with speed \cite{lopez86,zurbuchen02,bruno05,kasper12}.  Indeed, $\theta_{\alpha p}$ is positively correlated with wind speed \cite{marsch82a}.  In slower wind ($v_{rp} \lesssim 400\ \textrm{km/s}$), protons and $\alpha$-particles are often close to thermal equilibrium ($\theta_{\alpha p} \approx 1$), but in faster wind, protons and $\alpha$-particles usually have similar thermal speeds ($\theta_{\alpha p} \approx 4$) \cite{kasper08}.

Even so, deeper insight into the solar wind's non-thermal features (including $\theta_{\alpha p} \not= 1$) can be garnered from an analysis of \textit{collisional age}, $A_c$ \cite{feldman74,neugebauer76,marsch82a}.  The solar wind is affected by collisions among its constituent particles on a timescale, $\tau$, that varies with plasma conditions.  Collisional age is the number of such collisional timescales that elapse over the wind's expansion time:
\begin{equation} \label{eqn:ac:gen}
A_c \equiv \frac{r}{v_{rp}\,\tau}\,.
\end{equation}

The principal use of collisional age is the broad categorization of solar wind based on the overall progress of collisional thermalization: i.e., distinguishing \textit{collisionally young} ($A_c \ll 1$) and \textit{collisionally old} ($A_c \gg 1$) plasma.  Thus, while distinct definitions of $\tau$ exist for the various types of collisional interactions, a ``generic'' $\tau$ is often used in Equation \ref{eqn:ac:gen}.  One such definition is the \textit{self-collision time} \cite{book:spitzer56}, which, for a species $j$, is
\begin{equation} \label{eqn:tau:j}
\tau_j = \left( 11.4\,\frac{\textrm{s}}{\textrm{cm}^3\,\textrm{K}^{3/2}} \right) \left( \frac{T_j^{3/2}}{n_j} \right) \left( \frac{\mu_j^{1/2}}{Z_j^4} \right) \left( \frac{1}{\lambda_j} \right),
\end{equation}
where the Coulomb logarithm is
\begin{equation} \label{eqn:lambda}
\lambda_j = 9.42 + \ln\!\left[\left(\frac{1}{\textrm{cm}^{3/2}\,\textrm{K}^{3/2}}\right)\left(\frac{T_j^{3/2}}{n_j^{1/2}}\right)\left(\frac{1}{Z_j^2}\right)\right].
\end{equation}
In these equations, $\mu_j \equiv m_j\,/\,m_p$ and $Z_j \equiv \lvert q_j \rvert\,/\,q_p$, where $m_j$ and $q_j$ are, respectively, the mass and charge of a $j$-particle.  As protons are the most abundant ion species, their self-collision time can be used for collisional age.  Making the substitution $\tau=\tau_p$ into Equation \ref{eqn:ac:gen} gives
\begin{equation} \label{eqn:ac:spec}
A_c = \left(1.31\times10^7\,\frac{\textrm{cm}^3\,\textrm{km}\,\textrm{K}^{3/2}}{\textrm{s}\,\textrm{AU}}\right) \left(\frac{n_p}{v_{rp}\,T_p^{3/2}}\right) \left(r\right) \left(\lambda_p\right).
\end{equation}

Reference \cite{kasper08} qualitatively explored the influence of particle collisions on $\alpha$-proton thermalization by using observations from the \textit{Wind} spacecraft to plot the trend in $\theta_{\alpha p}$ versus solar wind speed and versus collisional age.  While $\theta_{\alpha p}$ was found to generally increase with speed, the trend exhibited considerable scatter (e.g., due to occasional fast wind with $\theta_{\alpha p} \approx 1$).  In contrast, the plot of $\theta_{\alpha p}$ versus collisional age showed a much tighter, smoother trend.  These results were interpreted as indicating that collisions strongly affect the $\alpha$-proton relative temperature in the solar wind.

\begin{table}
\caption{\label{tab:corr}Correlation Coefficients with $\theta_{\alpha p}$.}
\begin{ruledtabular}
\begin{tabular}{ld}
$x$       & \multicolumn{1}{c}{$\rho_S(x,\theta_{\alpha p})$} \\
\hline
$n_p$     & $-0.445$ \\
$v_{rp}$  & $0.607$  \\
$T_p$     & $0.737$  \\
$A_c$     & $-0.755$ \\
\end{tabular}
\end{ruledtabular}
\end{table}

The qualitative results of Reference \cite{kasper08} are confirmed by the quantitative results in Table \ref{tab:corr}, which lists the correlation coefficient, $\rho_S$, between each of four parameters ($n_p$, $v_{rp}$, $T_p$, and $A_c$) and $\theta_{\alpha p}$.  As indicated by the subscript ``$S$,'' these calculations used the Spearman correlation coefficient \cite{spearman04} rather than the more commonly-used Pearson correlation coefficient \cite{pearson96}.  Spearman's definition is less sensitive to outliers and is more general in that it gauges the monotonicity (versus the linearity) of the relationship between two parameters.

Table \ref{tab:corr} shows that, while $\theta_{\alpha p}$ is correlated with each of the four parameters, the trend is strongest with $A_c$.  Though the parameters $n_p$, $v_{rp}$, and $T_p$ are well known to themselves be correlated, $A_c$ combines them (see Equation \ref{eqn:ac:spec}) to produce a correlation with $\theta_{\alpha p}$ that is stronger than that with any one individually.  This result provides quantitative evidence that the $\theta_{\alpha p}$-values observed at $r = 1\ \textrm{AU}$ are heavily influenced by particle collisional -- more so even than differences between the processes that generate the slow and fast wind in the corona.

% Modeling collisional thermalization

While collisional age is a useful tool for broadly categorizing the collisionality of solar wind plasma, it has two significant limitations.  First, per Equation \ref{eqn:ac:gen}, collisional age is defined in terms of a collisional timescale.  To derive Equation \ref{eqn:ac:spec}, a ``generic'' timescale was chosen, but, as noted above, collisional relaxation occurs on different rates for different non-equilibrium features.  Second, Equation \ref{eqn:ac:gen} tacitly assumes that the parameters $n_p$, $v_{rp}$, and $T_p$ remain constant as the plasma travels from the Sun to the observer.  In reality, these parameters are affected by numerous processes (e.g., expansion and wave dissipation) and thus vary with solar distance, $r$.

As opposed to collisional age, a more complete understanding of how particle collisions impact the $\alpha$-proton relative temperature, $\theta_{\alpha p}$, can be achieved by directly modeling the collisional thermalization of these two species \cite{feldman74}.  Reference \cite[p.~34]{tech:huba11} considers a multi-species plasma with neither temperature anisotropy nor relative drift and analytically describes the time evolution of each species' temperature under the influence of particle collisions.  In particular,
\begin{equation} \label{eqn:dtdt} \begin{aligned}
\frac{dT_j}{dt} = \sum_{j'\not=j}& \left( 0.174\,\frac{\textrm{cm}^3\,\textrm{K}^{3/2}}{\textrm{s}} \right) \\
& \left( \frac{\left(\mu_j\,\mu_{j'}\right)^{1/2}Z_j^2\,Z_{j'}^2\,n_{j'}\,\lambda_{jj'}}{\left(\mu_j\,T_{j'}+\mu_{j'}\,T_j\right)^{3/2}} \right) \left( T_{j'} - T_j\right ),
\end{aligned} \end{equation}
where $j$ is a particle species in the plasma, the sum is taken over all other particle species $j'$ therein, and
\begin{equation} \label{eqn:lambda:jj} \begin{aligned}
\lambda_{jj'} = \lambda_{j'j} = 9. + \ln\!&\left[ \left(\frac{1}{\textrm{cm}^{3/2}\,\textrm{K}^{3/2}}\right) \left(\frac{Z_j\,Z_{j'}\left(\mu_j+\mu_{j'}\right)}{\mu_j\,T_{j'}+\mu_{j'}\,T_j}\right) \vphantom{\left(\frac{Z_j^2}{T_j}\right)^{1/2}} \right. \\
                               & \ \ \left. \left(\frac{n_j\,Z_j^2}{T_j} + \frac{n_{j'}\,Z_{j'}^2}{T_{j'}}\right)^{1/2} \right],
\end{aligned} \end{equation}
is the Coulomb logarithm.

For this study, Equation \ref{eqn:dtdt} was used to develop a simple model the radial evolution of $\theta_{\alpha p}$ in a parcel of solar wind plasma.  For this analysis, only protons and $\alpha$-particles were considered: other ions species and electrons were neglected.  An equation for $d\theta_{\alpha p}/dt$ was derived from Equation \ref{eqn:dtdt} using the chain rule.  By then assuming a system in steady state, the total derivative was converted into the convective derivative.  This readily gave
\begin{equation} \label{eqn:diffeq} \begin{aligned}
\frac{d\theta_{\alpha p}}{dr} =& \left(2.60\times10^7\,\frac{\textrm{cm}^3\,\textrm{km}\,\textrm{K}^{3/2}}{\textrm{s}\,\textrm{AU}}\right) \left(\frac{n_p}{v_{rp}\,T_p^{3/2}}\right) \\
                               & \left(\frac{\mu_\alpha^{1/2}\,Z_\alpha^2\left(1-\theta_{\alpha p}\right)\left(1+\eta_{\alpha p}\,\theta_{\alpha p}\right)}{\left(\mu_\alpha+\theta_{\alpha p}\right)^{3/2}}\right) \left(\lambda_{\alpha p}\right),
\end{aligned} \end{equation}
with the Coulomb logarithm
\begin{equation} \label{eqn:lambda:ap} \begin{aligned}
\lambda_{\alpha p} = 9. + \ln\!&\left[ \left(\frac{1}{\textrm{cm}^{3/2}\,\textrm{K}^{3/2}}\right) \left(\frac{T_p^{3/2}}{n_p^{1/2}}\right) \right. \\
                               & \ \ \left. \left(\frac{\mu_\alpha+\theta_{\alpha p}}{Z_\alpha\left(1+\mu_\alpha\right)}\right) \left(1+\frac{Z_\alpha^2\,\eta_{\alpha p}}{\theta_{\alpha p}}\right)^{-1/2} \right].
\end{aligned} \end{equation}
and $\eta_{\alpha p} \equiv n_\alpha\,/\,n_p$.  In the interest of generality, Equations \ref{eqn:diffeq} and \ref{eqn:lambda:ap} retain all factors of $\mu_\alpha$ and $Z_\alpha$.

Unlike in the definition of collisional age, the parameters $n_p$, $v_{rp}$, and $T_p$ in Equations \ref{eqn:diffeq} and \ref{eqn:lambda:ap} need not be constants but can instead vary with $r$.  This study assumed the following radial scalings:
\begin{equation} \label{eqn:scalings}
n_p(r) \propto r^{-1.8}\,,\ v_{rp}(r) \propto r^{-0.2}\,,\ \textrm{and}\ \ T_p(r) \propto r^{-0.74}\,.
\end{equation}
The scalings for $n_p$ and $T_p$ were derived from an analysis of observations from the \textit{Helios} spacecraft \cite{hellinger11}, and that for $v_{rp}$ was chosen to conserve proton flux density.  Some systematic effects inevitably resulted using these scaling as they are broad averages and are  partially coupled (e.g., the scalings of $n_p$ and $T_p$ vary with $v_{rp}$).  Nevertheless, the results presented below were found to be relatively insensitive to the specific scalings used.

Equation \ref{eqn:diffeq} (along with the Equations \ref{eqn:lambda:ap} and \ref{eqn:scalings}) was applied to each \textit{Wind} ion spectrum from the dataset.  More specifically, the set of observed $n_p$-, $v_{rp}$-, $T_p$-, $\eta_{\alpha p}$-, and $\theta_{\alpha p}$-values from each spectrum was used as a boundary condition at $r = 1.0\ \textrm{AU}$ in Equation \ref{eqn:diffeq}, which was then numerically solved so that the value of $\theta_{\alpha p}$ at some other $r$ could be inferred.  In these calculations, the impact of the singularity at $\theta_{\alpha p} = 1$ was mitigated by numerically integrating $\ln\!\left\lvert\theta_{\alpha p} - 1\right\rvert$ rather than $\theta_{\alpha p}$ per se.

\begin{figure}
\includegraphics[width=3in]{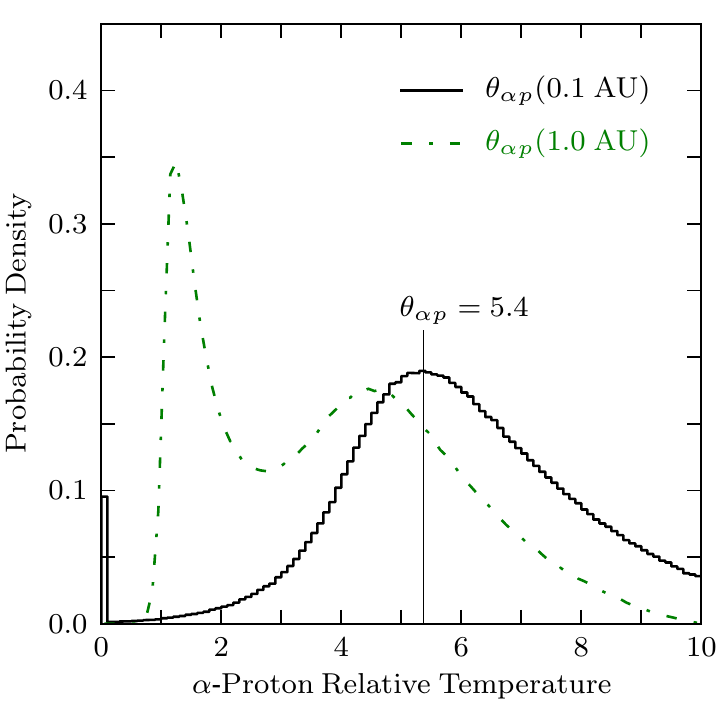}
\caption{\label{fig:comp}Distribution of $\theta_{\alpha p}$-values inferred for a distance $r=0.1\ \textrm{AU}$.  For reference, the distribution of $\theta_{\alpha p}$-values observed by the \textit{Wind} spacecraft at $r=1.0\ \textrm{AU}$ (see Figure \ref{fig:meas}) is also shown (green, dash-dotted curve) for ease of comparison.}
\end{figure}

The black, solid histogram in Figure \ref{fig:comp} shows the distribution of $\theta_{\alpha p}$-values computed in this way for $r = 0.1\ \textrm{AU} \approx 22 R_{\odot}$ (i.e., near the Alfv\'{e}n critical point).  Essentially, this is the distribution of $\alpha$-proton relative temperatures that is expected just outside the corona based on observations of solar wind near Earth.  Note that the narrow spike near $\theta_{\alpha p}(0.1\ \textrm{AU}) = 0$ is non-physical and was most likely caused by the singularity in Equation \ref{eqn:diffeq} and finite measurement uncertainty (see \cite{Note1}).

Statistically, these inferred $\theta_{\alpha p}(0.1\ \textrm{AU})$-values are most remarkable for having only a single mode.  While the measured values of $\theta_{\alpha p}(1.0\ \textrm{AU})$ have a bimodal distribution, Figure \ref{fig:comp} reveals that the distribution of the associated $\theta_{\alpha p}(0.1\ \textrm{AU})$-values has only one peak.  A Gaussian fit of this peak's crest indicates the mode of $\theta_{\alpha p}(0.1\ \textrm{AU})$ to be $5.4$.  Furthermore, this peak bears a striking resemblance in location, width, and shape to the peak near $\theta_{\alpha p}=4.5$ in the measured $\theta_{\alpha p}(1.0\ \textrm{AU})$-distribution.

These results, despite the simplicity of the analytic model used to obtain them, indicate that collisional thermalization, in and of itself, can account for the bimodality in the distribution of $\theta_{\alpha p}$-values observed at $r = 1.0\ \textrm{AU}$.  As stated above, the low-$\theta_{\alpha p}$ mode is predominantly associated with slow wind, and the high-$\theta_{\alpha p}$ mode predominately with fast wind.  Nevertheless, this correlation does not seem to arise from slow and fast wind having different coronal heating profiles.  Rather, slow wind simply has a longer expansion time and, being typically denser and cooler, thermalizes more rapidly (note the factor $n_p\,v_{rp}^{-1}\,T_p^{-3/2}$ in Equations \ref{eqn:ac:spec} and \ref{eqn:diffeq}).

Despite well-established differences in slow and fast wind at $r=1.0\ \textrm{AU}$, the results of this study suggest that such differences (at least in terms of relative ion temperatures) may be much less pronounced closer to the Sun.  Indeed, observations of coronal $\textrm{O}^{5+}$ with the \textit{Solar and Heliospheric Observatory}'s Ultraviolet Coronagraph Spectrometer have revealed evidence of enhanced heavy-ion temperatures both in sources of slow wind \cite{strachan02,frazin03} as well as in sources of fast wind \cite{kohl98}.  Likewise, other studies have found the energy flux density of the solar wind to be largely independent of wind speed \cite[and references therein]{lechat12}.  Collectively, these results suggest significant similarities in the mechanisms responsible for heating slow and fast wind in the solar corona.  The veracity of this conclusion may ultimately be evaluated with observations from \textit{Solar Probe Plus}, which is currently slated to have perihelia at $r<0.05\ \textrm{AU}$ \cite{guo10}.

%%% Acknowledgements

\textit{Acknowledgments}.  The authors offer their thanks to W.~H.~Matthaeus for fruitful discussions and to one of the anonymous referees for proposing the logarithmic integration of Equation \ref{eqn:diffeq}.  BAM is supported by the Charles Hard Townes Postdoctoral Fellowship.  BAM, SDB and LS-V acknowledge Marie Curie Project FP7 PIRSES-2010-269297--``Turboplasmas'' and NASA Contract Number NNN06AA01C.  This research has made use of the SAO/NASA Astrophysics Data System (ADS).

%%% Bibliography

%Merlin.mbs v4.21 2009-07-09.
%

\end{document}